\documentclass[prd,aps,reprint,noeprint,superscriptaddress,amsmath, amssymb]{revtex4-1}
\usepackage{graphicx}
\usepackage{subfigure}
\usepackage{lineno}
\usepackage{epsfig}
\usepackage{multirow}
\usepackage{overpic}
\usepackage{color}
\usepackage{bm} 
\usepackage{xspace} 
\usepackage{dcolumn}
\usepackage{booktabs,tabularx}
\usepackage{array}
\usepackage{textcomp}
\usepackage[colorlinks,linkcolor=blue,urlcolor=blue,citecolor=blue]{hyperref}
\usepackage{tikz}

\usepackage{cleveref}

\begin{document}
\normalsize
\parskip=5pt plus 1pt minus 1pt


\title{
\boldmath Strong and weak $CP$ tests in sequential decays of polarized $\Sigma^0$ hyperons
}

\author{
{\small M.~Ablikim$^{1}$, M.~N.~Achasov$^{4,c}$, P.~Adlarson$^{76}$, O.~Afedulidis$^{3}$, X.~C.~Ai$^{81}$, R.~Aliberti$^{35}$, A.~Amoroso$^{75A,75C}$, Q.~An$^{72,58,a}$, Y.~Bai$^{57}$, O.~Bakina$^{36}$
% ...
\\
\vspace{0.2cm}
(BESIII Collaboration)\\
\vspace{0.2cm} {\it
$^{1}$ Institute of High Energy Physics, Beijing 100049, People's Republic of China\\
% ...
$^{3}$ Bochum  Ruhr-University, D-44780 Bochum, Germany\\
$^{4}$ Budker Institute of Nuclear Physics SB RAS (BINP), Novosibirsk 630090, Russia\\
% ...
$^{35}$ Johannes Gutenberg University of Mainz, Johann-Joachim-Becher-Weg 45, D-55099 Mainz, Germany\\
$^{36}$ Joint Institute for Nuclear Research, 141980 Dubna, Moscow region, Russia\\
% ...
$^{57}$ Southeast University, Nanjing 211100, People's Republic of China\\
$^{58}$ State Key Laboratory of Particle Detection and Electronics, Beijing 100049, Hefei 230026, People's Republic of China\\
% ...
$^{72}$ University of Science and Technology of China, Hefei 230026, People's Republic of China\\
% ...
$^{75}$ University of Turin and INFN, (A)University of Turin, I-10125, Turin, Italy; (B)University of Eastern Piedmont, I-15121, Alessandria, Italy; (C)INFN, I-10125, Turin, Italy\\
$^{76}$ Uppsala University, Box 516, SE-75120 Uppsala, Sweden\\
% ...
$^{81}$ Zhengzhou University, Zhengzhou 450001, People's Republic of China\\
\vspace{0.2cm}
$^{a}$ Deceased\\
% ...
$^{c}$ Also at the Novosibirsk State University, Novosibirsk, 630090, Russia\\
% ...
}\vspace{0.4cm}  }  
} 
\begin{abstract}
    The $J/\psi, \psi(3686) \to \Sigma^0 \bar{\Sigma}^{0}$ processes and
    subsequent decays are studied using the world's largest $J/\psi$ and
    $\psi(3686)$ data samples collected with the BESIII detector.  
    The parity-violating decay parameters of the decays $\Sigma^0 \to \Lambda \gamma$ 
    and $\bar{\Sigma}^0 \to \bar{\Lambda} \gamma$, 
    $\alpha_{\Sigma^0} = -0.0017 \pm 0.0021 \pm 0.0018$ and
    $\bar{\alpha}_{\Sigma^0} = 0.0021 \pm 0.0020 \pm 0.0022$, are measured 
    for the first time.  
    The strong-$CP$ symmetry is tested in the decays of the $\Sigma^0$ 
    hyperons for the first time by measuring the asymmetry 
    $A^{\Sigma}_{CP}=\alpha_{\Sigma^0} + \bar{\alpha}_{\Sigma^0} = (0.4 \pm 2.9 \pm 1.3)\times 10^{-3}$.
    The weak-$CP$ test is performed in the subsequent decays of their daughter
    particles $\Lambda$ and $\bar{\Lambda}$. Also for the first time, the
    transverse polarizations of the $\Sigma^0$ hyperons in $J/\psi$ and
    $\psi(3686)$ decays are observed with opposite directions, and the
    ratios between the S-wave and D-wave contributions of the $J/\psi,
    \psi(3686) \to \Sigma^0 \bar{\Sigma}^{0}$ decays are obtained. These
    results are crucial to understand the decay dynamics of the charmonium
    states and the production mechanism of the $\Sigma^0-\bar{\Sigma}^0$
    pairs.
\end{abstract}

\maketitle


The question of why our universe consists of matter, but almost no
antimatter, has puzzled the scientific community for more than half a
century and is to this day subjected to intensive
research~\cite{BESIII:2021ypr, BESIII:2022qax, BESIII:2023drj,
  LHCb:2023zcp, Belle:2024dhj, LHCb:2023rae}.  One of the
long-standing explanations is baryogenesis~\cite{Sakharov:1967dj},
i.e. that the matter abundance is generated dynamically.  Among the
necessary criteria for this to be possible is the existence of
processes that violate charge-conjugation and parity ($CP$)
conservation.  The Standard Model (SM) of particle physics allows for
tiny signals of $CP$ violation ($CPV$) in weak interactions, in line
with experimental observations in meson
decays~\cite{Christenson:1964fg, Aubert:2001nu, Abe:2001xe,
  Aaij:2019kcg}.  The SM should in principle also allow for $CP$
violating strong processes, which would be manifest in e.g.  a
non-zero neutron electric dipole moment (EDM)~\cite{Ramsey:1982pq,
  Fortson:2003fi, Luders:1957bpq}.  However, the extremely tiny
experimental upper limit of the neutron EDM~\cite{Abel:2020pzs}
suggests that $CPV$ in strong interactions is unnaturally small.  The
SM itself does not provide an answer to the smallness of strong
$CPV$, which calls for further investigations.  The radiative decay
$\Sigma^0 \to \Lambda \gamma$ could in principle offer an opportunity
to study the interference between a parity-conserving (from the
magnetic transition moment) and a parity-violating amplitude (from the
electric dipole transition moment, EDTM). The EDTM of this decay is
related to the neutron EDM via SU(3) flavor
symmetry~\cite{Gell-Mann:1961omu, Nair:2018mwa}. 
This can be explored by measuring the parity-violating decay parameters 
$\alpha_{\Sigma^0}$, and $\bar{\alpha}_{\Sigma^0}$ for the charge conjugate decay. 
In the SM, $\alpha_{\Sigma^0}$ and $\bar{\alpha}_{\Sigma^0}$ are predicted to be very small~\cite{Nair:2018mwa},
nonzero measurements of these parameters at current or near-future experimental sensitivity 
levels would be indications of parity violation beyond the standard model.
Based on the values of these parameters, the strong-$CP$ symmetry can be tested by
measuring the asymmetry $A^{\Sigma}_{CP} = \alpha_{\Sigma^0} + \bar{\alpha}_{\Sigma^0}$,
finding a non-vanishing $A^{\Sigma}_{CP}$ would point to signals of $CPV$ 
beyond the SM~\cite{Nair:2018mwa}.

The large yield of quantum entangled $\Sigma^0 \bar{\Sigma}^0$ pairs
at BESIII~\cite{Li:2016tlt} enables a pioneering test of
strong-$CP$ symmetry in the $\Sigma^0$ hyperon decays.  In addition,
the subsequent decays of the daughter particles $\Lambda$ and
$\bar{\Lambda}$ provide an independent measurement of the
$\Lambda/\bar{\Lambda}$ decay parameter
$\alpha_{\Lambda}/\bar{\alpha}_{\Lambda}$, which have undergone much
scrutiny in recent years due to the large discrepancy observed between
old and new data \cite{BESIII:2018cnd, BESIII:2022qax, BESIII:2021ypr,
  BESIII:2023drj, Astbury:1975hn, Cleland:1972fa, Overseth:1967zz,
  Dauber:1969hg}, and allow a test of weak-$CP$ symmetry.  Finally, the
data collected with different vector charmonia will help elucidate their
decay dynamics.  

The decay of a charmonium ($\psi$) into a hyperon-antihyperon pair can
be completely described by two parameters, $\alpha_{\psi}$ and $\Delta
\Phi_{\psi}$~\cite{Faldt:2017kgy}.  The former describes the angular
distribution of the hyperon-antihyperon pair, while the latter is
related to the hyperon polarization,
\begin{equation}
    \label{eq:polarization}
    P_y(\cos \theta_{Y}) = \frac{\sqrt{1-\alpha^2_{\psi}} \sin\Delta\Phi_{\psi} \cos \theta_{Y} \sin\theta_{Y}}{1+\alpha_{\psi}\cos^2\theta_{Y}},
\end{equation}
where $\theta_{Y}$ is the angle between the momenta of $e^+$ and the
hyperon in the $e^+ e^-$ center-of-mass (C.M.) system.  The hyperon
polarization would manifest itself by a non-zero $\Delta \Phi_{\psi}$.

In recent years, the BESIII collaboration has published a series of
studies~\cite{BESIII:2018cnd, BESIII:2020fqg, BESIII:2020lkm,
  BESIII:2021ypr, BESIII:2022qax, BESIII:2023drj, BESIII:2022lsz,
  BESIII:2023lkg, BESIII:2023euh} about hyperon polarization.  An
intriguing result was reported in Ref.~\cite{BESIII:2020fqg}, where
the directions of the $\Sigma^+/\bar{\Sigma}^-$ polarizations were
observed to be opposite in $J/\psi$ and $\psi(3686)$ decays; this
phenomenon, however, is not observed in $J/\psi \to \Xi^{-}
\bar{\Xi}^+ / \Xi^0 \bar{\Xi}^0$ and $\psi(3686) \to \Xi^{-}
\bar{\Xi}^+ / \Xi^0 \bar{\Xi}^0$ decays~\cite{BESIII:2021ypr,
  BESIII:2022lsz, BESIII:2023lkg, BESIII:2023drj, BESIII:2021ypr}.
Until now, there is no interpretation of these results, therefore,
more experimental measurements of the hyperon polarization are highly
desirable.

The analysis presented in this Letter is based on samples of
$(10087 \pm 44) \times 10^6$ $J/\psi$ and $(2712 \pm 14) \times 10^6$
$\psi(3686)$ events~\cite{BESIII:2021cxx, BESIII:2024lks} collected
with the BESIII detector at the BEPCII collider.  Details about BEPCII
and BESIII can be found in Refs.~\cite{Ablikim2010, BESIII:2020nme,
  Yu:2016cof, Huang:2022wuo, Li:2017eToF, Guo:2017eToF, Cao:2020ibk}.
Simulated data samples produced with  {\sc
  geant4}-based~\cite{G42002iii} Monte Carlo (MC) software, which
includes the geometric description of the BESIII detector and the
detector response, are used to determine detection efficiencies and
estimate backgrounds.  The simulations model the beam energy spread
and initial state radiation (ISR) in the $e^+e^-$ annihilations with
the generator {\sc kkmc}~\cite{Jadach2001}.  Inclusive MC samples
of $J/\psi$ and $\psi(3686)$ resonances are produced, in which the
known decay modes are modeled with {\sc
  evtgen}~\cite{Lange2001,*Ping2008}, and the remaining unknown decays
  are modeled with {\sc lundcharm}~\cite{Chen2000,*Yang2014a}.  For
  the signal process, $e^+ e^- \to \psi \to \Sigma^{0} \bar{\Sigma}^0,
  \Sigma^{0} \to \Lambda (\to p \pi^-) \gamma,\bar{\Sigma}^0 \to
  \bar{\Lambda} (\to \bar{p}\pi^+) \gamma$, two different MC samples
  are used.  One is generated,  
  with the uniform distribution of the kinematic momenta of the final states,
  to obtain the efficiency correction used in the fit to determine the
  parameters, and the other is generated according to the joint
  angular distribution for the signal process with the parameters
  obtained by this analysis (signal MC).

The $\Lambda$ and $\bar{\Lambda}$ hyperons
are reconstructed from their dominant decay modes, $\Lambda \to p\pi^-$
and $\bar{\Lambda} \to \bar{p} \pi^+$.  Charged tracks detected in the
multilayer drift chamber (MDC) are required to be within a polar angle
($\theta$) range of $|\rm{cos\theta}|<0.93$, where $\theta$ is defined
with respect to the $z$-axis, which is the symmetry axis of the MDC.
Events with at least four charged tracks are retained.  Tracks with
momentum larger than $0.45$~GeV/$c$ for the $J/\psi$ dataset and
$0.55$~GeV/$c$ for the $\psi(3686)$ dataset are considered as proton
candidates, otherwise as pion candidates.  There is no further
particle identification requirement.  Secondary vertex
fits~\cite{Xu:2009zzg} are performed for all combinations with
oppositely-charged proton and pion candidates, and the combinations
passing the vertex fit successfully and with invariant masses within a
range of $[1.111, 1.120]$~GeV/$c^2$ are regarded as the $\Lambda$ 
and $\bar{\Lambda}$ candidates. 

Photon candidates are identified using showers in the electromagnetic
calorimeter~(EMC).  The deposited energy of each shower must be more
than 25~MeV in the barrel region($|\cos\theta| < 0.80$) and more than
50~MeV in the end-cap region ($0.86<|\cos\theta|<0.92$).  To suppress
electronic noise and showers unrelated to the event, the difference
between the EMC time of the photon candidate and the event start time
is required to be within [0, 700]~ns.

Events with at least one $\Lambda$ candidate, one $\bar{\Lambda}$
candidate, and two photons are considered for further analysis.  A
four-constraint (4C) kinematic fit is performed under the $\Lambda
\bar{\Lambda} \gamma \gamma$ hypothesis by constraining their total
reconstructed four-momentum to that of the initial $\psi$. If there is
more than one $\Lambda \bar{\Lambda} \gamma \gamma$ combination, the
one with the smallest $\chi^2$ of the 4C fit ($\chi^2_{4 \rm C}$) is
selected, and $\chi^2_{4 \rm C}< 100$ is required. Since there are two
$\gamma$ candidates ($\gamma_1$, $\gamma_2$) per event, there will be
two combinations of $\Lambda(\bar{\Lambda})$ and $\gamma$.  Therefore,
a six-constraint (6C) kinematic fit is performed under the $\Lambda
\bar{\Lambda} \gamma \gamma$ hypothesis by additionally constraining
the invariant masses of $\Lambda \gamma_1$ $(m_{\Lambda \gamma_1})$ and
$\bar{\Lambda} \gamma_2$ $(m_{\bar{\Lambda} \gamma_2})$ to the known
masses~\cite{ParticleDataGroup:2022pth} of $\Sigma^0$ and
$\bar{\Sigma}^0$.  The combination which minimizes $\chi^2$ of the 6C
fit ($\chi^2_{6C}$) is kept, and there is no further requirement on the $\chi^2_{6C}$.
Finally, the $m_{\Lambda \gamma_1}$ and $m_{\bar{\Lambda} \gamma_2}$
masses from the 4C fit are required to be within $[1.178, 1.206]$~GeV/$c^2$.
After applying all the event selection criteria,
the final event samples in the signal region contain  
$1083676$ events for the $J/\psi$ decay
and $51837$ events for the $\psi(3686)$ decay.

The backgrounds in this analysis are divided into three categories:
background channels which have a $\Sigma^0 \bar{\Sigma}^0$ pair (type
1), channels that have one $\Sigma^0$ or one $\bar{\Sigma}^0$ (type
2), and ones containing no $\Sigma^0$ or $\bar{\Sigma}^0$ (type 3).
Inclusive MC samples of 10 billion $J/\psi$ events and 2.7 billion
$\psi(3686)$ events are used for studying background
type 1.  After applying the same selection criteria as for data, the
only peaking background channel is found to be $J/\psi \to \gamma
\eta_c, \eta_c \to \Sigma^0 \bar{\Sigma}^0$.  An exclusive MC
simulation of this process is carried out, and the corresponding
number of events from this channel in data is estimated to be
$391\pm145$, which is only about 0.036\% of the total selected
candidates in the $J/\psi$ data sample and can thus be neglected.  The
later two types of backgrounds are estimated with the two-dimensional
sideband regions of the distribution of $m_{\bar{\Lambda} \gamma_2}$
versus $m_{\Lambda \gamma_1}$ using the same method as in
Ref.~\cite{BESIII:2020fqg}.  The lower and upper sideband regions are
defined as $m_{\Lambda\gamma_1 / \bar{\Lambda}\gamma_2} \in (1.138,
1.166)$~GeV/$c^2$ and $m_{\Lambda\gamma_1 / \bar{\Lambda}\gamma_2} \in
(1.218, 1.246)$~GeV/$c^2$, respectively.  The background levels are
found to be 0.7\% for $J/\psi \to \Sigma^0 \bar{\Sigma}^0$ and 1.5\%
for $\psi(3686) \to \Sigma^0 \bar{\Sigma}^0$. 

\begin{figure}[htbp]
    \begin{center}
        \begin{tikzpicture}[scale=1.0]
            \node(a) at (-1.5,0.0)
            {\includegraphics[width=0.9\linewidth]{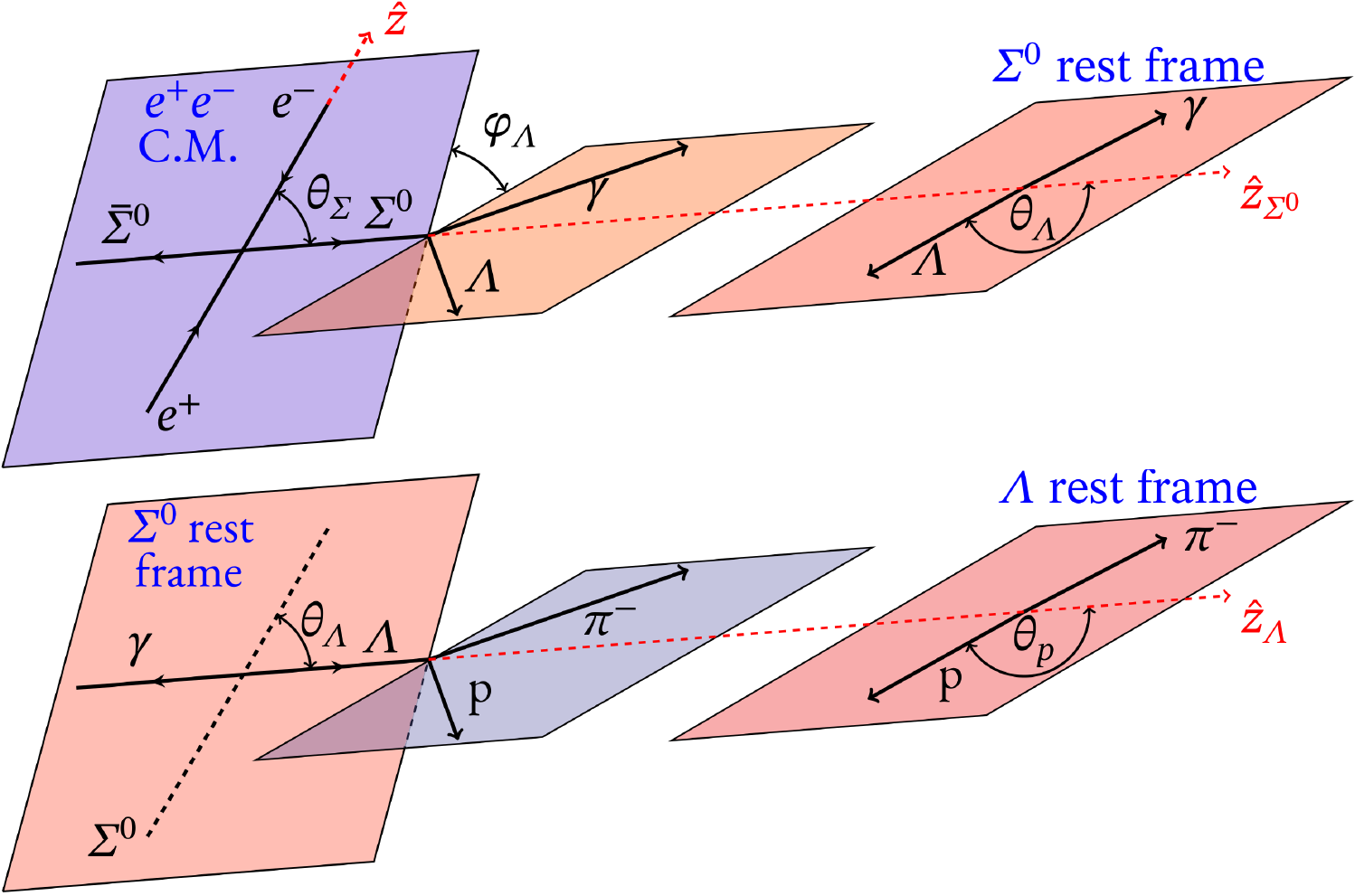}};
        \end{tikzpicture}
        \caption{The definitions of the helicity angles. The polar
          angle $\theta_{\Sigma}$ is the angle between the $\Sigma^0$ momentum
          and the $e^+$ beam direction in the $e^+ e^-$ C.M., 
          where the $\hat{z}$ axis is defined along the
          $e^+$ momentum.  $\theta_{\Lambda}$ and $\varphi_{\Lambda}$
          are the polar and azimuthal angles of the $\Lambda$ momentum
          direction in the $\Sigma^0$ rest frame, where $\hat{z}_{\Sigma^0}$
          is defined along the $\Sigma^0$ momentum direction in the $e^+
          e^-$ C.M., and $\hat{y}_{\Sigma^0}$ is defined by $\hat{z}
          \times \hat{z}_{\Sigma^0}$.  The angles $\theta_{p}$ and
          $\varphi_{p}$ are the polar and azimuthal angles of the proton
          momentum direction in the $\Lambda$ rest frame, where
          $\hat{z}_{\Lambda}$ is defined along the $\Lambda$ momentum
          in the $\Sigma^0$ rest frame, and $\hat{y}_{\Lambda}$ is
          along $\hat{z}_{\Sigma^0} \times \hat{z}_{\Lambda}$.  }
        \label{DecayPlane}
    \end{center}
\end{figure}

Following the formulation in Refs.~\cite{Dubnickova:1992ii,
  Gakh:2005hh, Czyz:2007wi, Commins:1983ns, Nair:2018mwa, Li:2019knv},
the joint angular distribution of the full decay chain $e^+ e^- \to
\psi \to \Sigma^{0} \bar{\Sigma}^0, \Sigma^{0} \to \Lambda (\to p
\pi^-) \gamma,\bar{\Sigma}^0 \to \bar{\Lambda} (\to \bar{p}\pi^+)
\gamma$ is obtained,

\vspace{-0.5cm}

\begin{align}
    \label{formulation}
    & \mathcal{W}(\vec{\zeta}, \vec{\omega})
    \propto (1-\alpha_{\Lambda}\alpha_{\Sigma^0}\cos\theta_p)(1-\bar{\alpha}_{\Lambda}\bar{\alpha}_{\Sigma^0}\cos\theta_{\bar{p}})\times \notag \\
    & \{1 + \alpha_{\psi}\cos^2\theta_{\Sigma} \notag + \sqrt{1-\alpha^2_{\psi}}\sin\theta_{\Sigma} \cos\theta_{\Sigma}  
    \cdot \notag \\
    & [\beta_{\gamma}\bar{\beta}_{\gamma} F_1 - (\beta_{\gamma} F_2 - \bar{\beta}_{\gamma} F_3)] + \notag \\
    & \beta_{\gamma}\bar{\beta}_{\gamma} [\alpha_{\psi} F_4 + F_5 - 
    (\alpha_{\psi} + \cos^2 \theta_{\Sigma})\cos\theta_{\Lambda}\cos\theta_{\bar{\Lambda}}]\}.
\end{align}
Here, $\vec{\omega}$ represents the six parameters of interest,
$\alpha_{\psi}$, $\Delta \Phi_{\psi}$, $\alpha_{\Sigma^0}$,
$\bar{\alpha}_{\Sigma^0}$, $\alpha_{\Lambda}$,
$\bar{\alpha}_{\Lambda}$; $\vec{\zeta}$ stands for seven helicity
angles, $\theta_{\Sigma}$, $\theta_{\Lambda}$, $\varphi_{\Lambda}$,
$\theta_{\bar{\Lambda}}$, $\varphi_{\bar{\Lambda}}$, $\theta_{p}$, and
$\theta_{\bar{p}}$.  For simplicity, the symbols $\beta_{\gamma}
\equiv \frac{\alpha_{\Sigma^0} - \alpha_{\Lambda}
  \cos\theta_{p}}{1-\alpha_{\Lambda}\alpha_{\Sigma^0}\cos\theta_{p}}$
and $\bar{\beta}_{\gamma} \equiv \frac{\bar{\alpha}_{\Sigma^0} -
  \bar{\alpha}_{\Lambda}
  \cos\theta_{\bar{p}}}{1-\bar{\alpha}_{\Lambda}
  \bar{\alpha}_{\Sigma^0}\cos\theta_{\bar{p}}}$ are introduced, and
$F_1-F_5$ are defined below, 

\vspace{-0.5cm}
\begin{align}
    \label{eq:F}
    & F_1 = \cos\Delta\Phi_{\psi} (\sin\theta_{\Lambda}\cos\varphi_{\Lambda}\cos\theta_{\bar{\Lambda}} -\sin\theta_{\bar{\Lambda}}\cos\varphi_{\bar{\Lambda}}\cos\theta_{\Lambda}), \notag \\
    & F_2 =  \sin \Delta\Phi_{\psi} \sin\theta_{\Lambda}\sin\varphi_{\Lambda}, \notag \\
    & F_3 = \sin \Delta\Phi_{\psi} \sin\theta_{\bar{\Lambda}}\sin\varphi_{\bar{\Lambda}}, \notag\\
    & F_4 = \sin^2\theta_{\Sigma} \sin \theta_{\Lambda}\sin\theta_{\bar{\Lambda}}\sin\varphi_{\Lambda}\sin\varphi_{\bar{\Lambda}}, \notag\\
    & F_5 = \sin^2\theta_{\Sigma} \sin \theta_{\Lambda}\sin\theta_{\bar{\Lambda}} \cos\varphi_{\Lambda}\cos\varphi_{\bar{\Lambda}}.
\end{align}
The helicity angles are
constructed as illustrated in Fig.~\ref{DecayPlane}, and the
corresponding angles of the anti-particle decay sequence are obtained
analogously.  To validate the formula, a check is made by setting the
parameters $\alpha_{\Sigma^0}$ and $\bar{\alpha}_{\Sigma^0}$ in
$\mathcal{W}(\vec{\zeta}, \vec{\omega})$ to zero, and then the
formalism is found to be equivalent to the covariant formalism given
in Ref.~\cite{Faldt:2019zdl}.

An unbinned maximum likelihood fit is performed in the seven angular
dimensions $\vec{\zeta}$ by simultaneously fitting both the $J/\psi
\to \Sigma^0 \bar{\Sigma}^0$ and $\psi(3686) \to \Sigma^0
\bar{\Sigma}^0$ datasets to determine the parameters $\vec{\omega}$.
During the fitting process, the background events
estimated from the data sidebands are included, and their 
log-likelihood is subtracted from the data based on the normalized weights.
The numerical results are summarized in Table~\ref{tab:final_results},
together with the $CP$ asymmetries $A^{\Sigma}_{CP}$ and $A^{\Lambda}_{CP}
= (\alpha_{\Lambda} + \bar{\alpha}_{\Lambda})/(\alpha_{\Lambda} -
\bar{\alpha}_{\Lambda})$. 
The correlation coefficient matrix of the fitted parameters is shown in Tab.~\ref{tab:correlation}.

\begin{table*}[!htbp]
\renewcommand\arraystretch{1.2}
    \caption{Full correlation coefficient matrix of the fitted parameters.
    }
\label{tab:correlation}
\centering
\scalebox{1.2}
{
    \begin{tabular}{l c c c c c c c c }
        \hline \hline 
        \noalign{\vskip 3pt}
        Coefficients    & $\alpha_{J/\psi}$ & $\Delta\Phi_{J/\psi}$ & $\alpha_{\psi(3686)}$ & $\Delta\Phi_{\psi(3686)}$ & $\alpha_{\Sigma^0}$ & $\bar{\alpha}_{\Sigma^0}$ & $\alpha_{\Lambda}$ & $\bar{\alpha}_{\Lambda}$ \\
        \hline
        $\alpha_{J/\psi}$           & 1.000  & -0.016 & -0.012 & 0.012  & -0.002 & -0.002 & 0.018  & -0.013 \\
        $\Delta\Phi_{J/\psi}$       & -0.016 & 1.000  & 0.006  & -0.007 & 0.004  & 0.000  & 0.011  & 0.027  \\
        $\alpha_{\psi(3686)}$       & -0.012 & 0.006  & 1.000  & -0.451 & 0.007  & -0.007 & -0.011 & 0.004  \\
        $\Delta\Phi_{\psi(3686)}$   & 0.012  & -0.007 & -0.451 & 1.000  & -0.001 & -0.001 & -0.051 & -0.067 \\
        $\alpha_{\Sigma^0}$         & -0.002 & 0.004  & 0.007  & -0.001 & 1.000  & -0.009 & -0.083 & -0.083 \\
        $\bar{\alpha}_{\Sigma^0}$   & -0.002 & 0.000  & -0.007 & -0.001 & -0.009 & 1.000  & 0.032  & 0.033  \\
        $\alpha_{\Lambda}$          & 0.018  & 0.011  & -0.011 & -0.051 & -0.083 & 0.032  & 1.000  & 0.980  \\
        $\bar{\alpha}_{\Lambda}$    & -0.013 & 0.027  & 0.004  & -0.067 & -0.083 & 0.033  & 0.980  & 1.000  \\
        \hline \hline
    \end{tabular}
    }
\end{table*}

\begin{table}[htbp]
\renewcommand\arraystretch{1.2}
    \caption{ Values and uncertainties of the fit parameters and the
      $CP$ asymmetries $A^{\Sigma}_{CP}$ and $A^{\Lambda}_{CP}$, along with the previous
      measurements. The first and second uncertainties in this work
      are statistical and systematic, respectively, and those for the
      previous results are the total uncertainties. 
      In the previous results, only the values reported in Ref.~\cite{BESIII:2017kqw} 
      are cited for $\alpha_{J/\psi}$ and $\alpha_{\psi(3686)}$, as it
      gives the most precise measurement for $\alpha_{J/\psi}$ and the only measurement for $\alpha_{\psi(3686)}$.
      }
\label{tab:final_results}
\centering
\scalebox{0.9}
{
    \begin{tabular}{l c c c}
        \hline \hline 
        \noalign{\vskip 3pt}
        Parameter                       & This work                                                    & Previous results\\
        \hline
        $\alpha_{J/\psi}$               & $-0.4133 \pm 0.0035 \pm  0.0077$                             &  $-0.449\pm0.022$~\cite{BESIII:2017kqw}\\
        $\Delta\Phi_{J/\psi}$(rad)      & $-0.0828 \pm 0.0068 \pm  0.0033$                             & ...    \\
        $\alpha_{\psi(3686)}$           & $0.814 \pm 0.028 \pm 0.028$                                  &  $0.71\pm0.12$~\cite{BESIII:2017kqw}\\
        $\Delta\Phi_{\psi(3686)}$(rad)  & $0.512 \pm 0.085 \pm 0.034$                                  & ...    \\
        $\alpha_{\Sigma^0}$               & $-0.0017 \pm 0.0021 \pm 0.0018$                              & ...    \\
        $\bar{\alpha}_{\Sigma^0}$         & ${\color{white}-}0.0021 \pm 0.0020 \pm 0.0022$               & ...    \\
        $\alpha_{\Lambda}$              & ${\color{white}-}0.730 \pm 0.051 \pm 0.011$                  & ${\color{white}-}0.748\pm 0.007$~\cite{ParticleDataGroup:2022pth}        \\
        $\bar{\alpha}_{\Lambda}$        & $-0.776 \pm 0.054 \pm 0.010$                                 & $-0.757\pm0.004$~\cite{ParticleDataGroup:2022pth}        \\
        \hline 
        $A^{\Sigma}_{CP}$               & ${\color{white}-}(0.4 \pm 2.9 \pm 1.3) \times 10^{-3}$       & ...    \\
        $A^{\Lambda}_{CP}$              & $(-3.0 \pm 6.9 \pm 1.5) \times 10^{-2}$                      & $(-2.5 \pm 4.8)\times 10^{-3}$~\cite{BESIII:2022qax} \\  
        \hline \hline
    \end{tabular}
    }
\end{table}

\begin{figure}[htbp]
    \begin{center}
        \mbox{
            \put(-120, 0){
                \begin{overpic}[width = 0.9\linewidth]{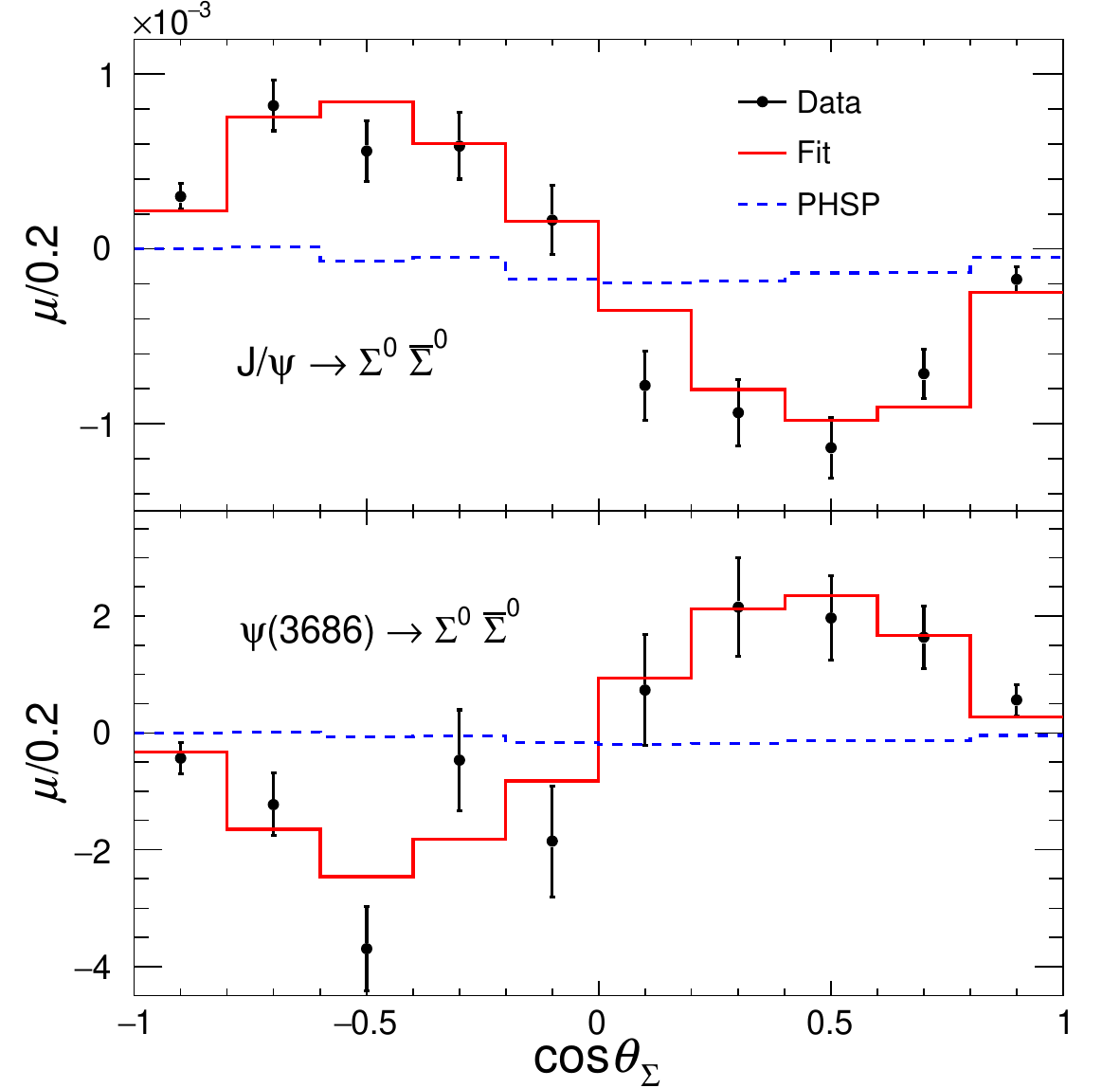}
                \end{overpic}
            }
        }
    \end{center}
    \vspace{-0.5cm}
    \caption{Distributions of the moments $\mu$ versus
      $\cos\theta_{\Sigma}$ for $J/\psi \to \Sigma^0 \bar{\Sigma}^0$
      and $\psi(3686) \to \Sigma^0 \bar{\Sigma}^0$.  The points with
      error bars are data, the red solid lines are the signal MC
      samples with input parameters fixed to the fit results, and the
      blue dashed lines represent the distributions without
      polarization from the PHSP MC samples.}
    \label{Polarization}
\end{figure}

The transverse polarizations (parallel or anti-parallel to
$\hat{y}_{\Sigma^0}$) of the $\Sigma^0$ hyperon in the $J/\psi$ and
$\psi(3686)$ decays are observed for the first time and are found to
have opposite directions.  The corresponding parameters are determined
to be $\Delta \Phi_{J/\psi} = -0.0828 \pm 0.0068 \pm 0.0033$~rad and
$\Delta \Phi_{\psi(3686)} = 0.512 \pm 0.085 \pm 0.035$~rad for the
$J/\psi \to \Sigma^0 \bar{\Sigma}^0$ and $\psi(3686) \to \Sigma^0
\bar{\Sigma}^0$ decays, respectively, which differ from zero with a
significance of $12.2\sigma$ for the $J/\psi$ dataset and $7.4\sigma$
for the $\psi(3686)$ dataset.  The $\Sigma^0$ polarizations can be
illustrated through the moment $\mu$, as defined:

\vspace{-0.5cm}
\begin{align}
    \label{eq:mu}
    \mu^k(\cos\theta_{\Sigma}) & =   \notag \\
    \frac{1}{N_{\rm total}} \sum^{N^k}_{i} (\sin\theta^i_{\Lambda} \sin\varphi^i_{\Lambda} \cos\theta^i_{p} & + \sin\theta^i_{\bar{\Lambda}} \sin\varphi^i_{\bar{\Lambda}} \cos\theta^i_{\bar{p}}),
\end{align}
where $N_{\rm total}$ is the total number of events in the dataset,
$N^k$ is the number of events in the $k$-th $\cos\theta_{\Sigma}$ bin,
and $i$ is the $i$-th event in that bin.  The expected angular
dependence of the moment for the acceptance-corrected data is
$(\alpha_{\Lambda} - \bar{\alpha}_{\Lambda}) (1+\alpha_{\psi} \cos^2
\theta_{\Sigma}) P_y(\cos \theta_{\Sigma})/(18+6\alpha_{\psi})$.
Comparing the data to the PHSP MC sample, as shown in
Fig.~\ref{Polarization}, the polarizations of the $\Sigma^0$ are
observed clearly.

\begin{figure}[htbp]
    \begin{center}
        \mbox{
            \put(-125, 0){
                \begin{overpic}[width = 0.95\linewidth]{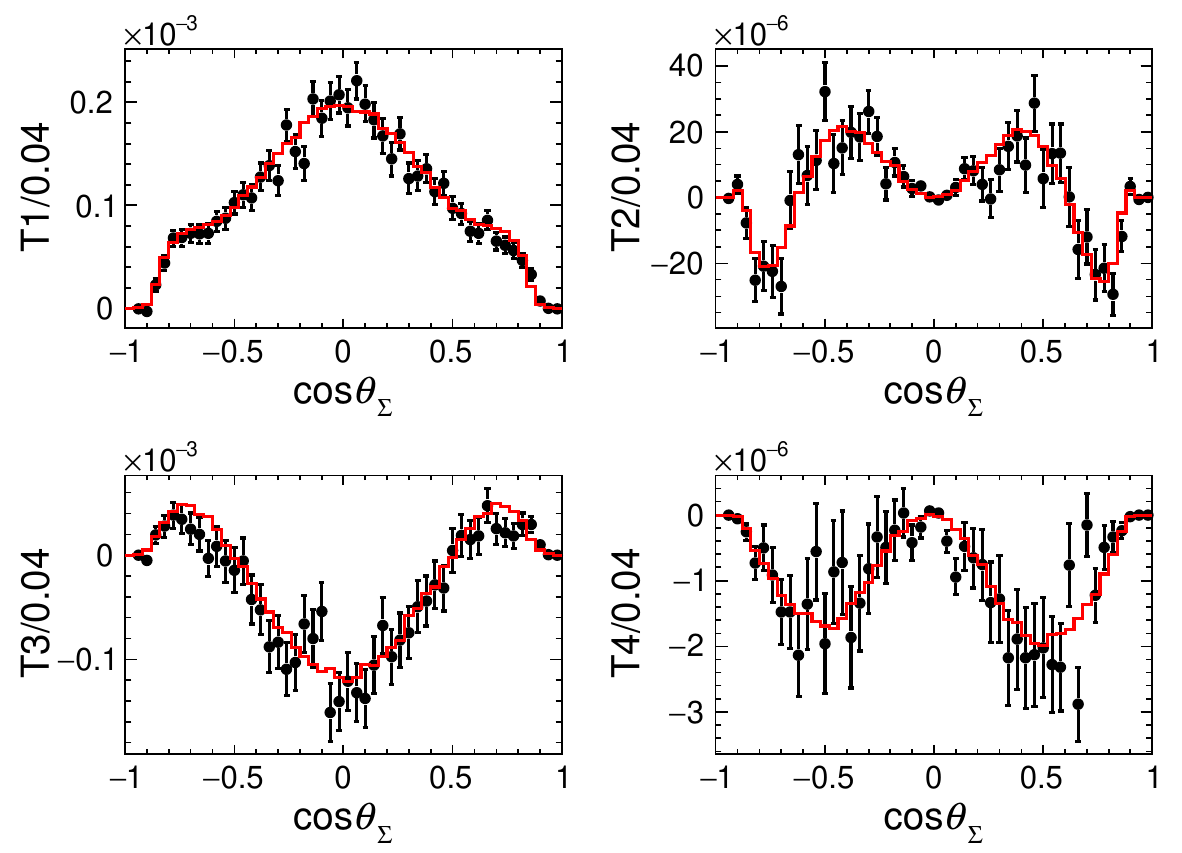}
                \end{overpic}
            }
        }
    \end{center}
    \vspace{-0.5cm}
    \caption{Distributions of the moments $\rm T1-T4$ versus $\cos
      \theta_{\Sigma}$ for the $J/\psi$ dataset. The dots with error bars represent the data, 
      and the red solid lines are the signal MC samples
      with the input parameters fixed to the fit results.}
    \label{T1_T4}
\end{figure}

\begin{figure}[htbp]
    \begin{center}
        \mbox{
            \put(-125, 0){
                \begin{overpic}[width = 0.95\linewidth]{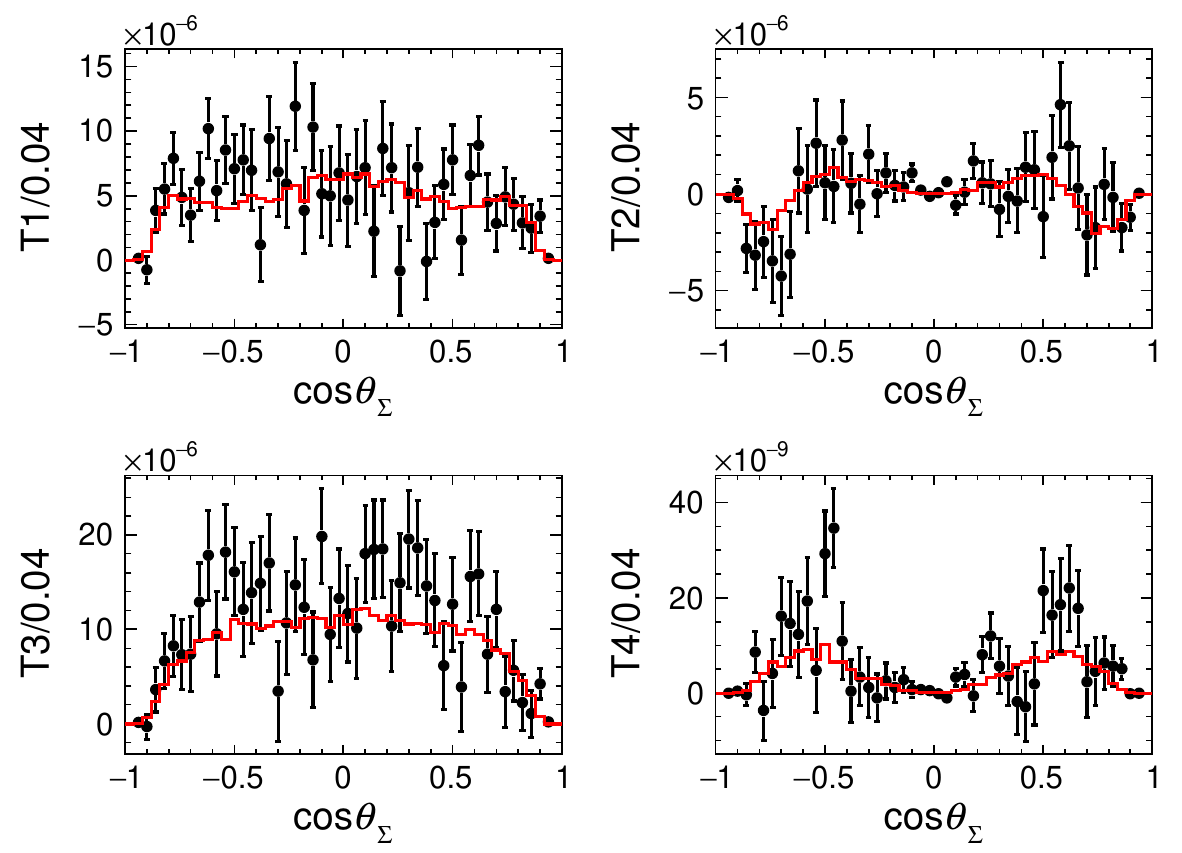}
                \end{overpic}
            }
        }
    \end{center}
    \caption{Distributions of the moments $\rm T1-T4$ versus $\cos
      \theta_{\Sigma}$ for the $\psi(3686)$ dataset. The dots with error bars represent the data, 
      and the red solid lines are the signal MC samples
      with the input parameters fixed to the fit results.}
    \label{T1_T4_3686}
\end{figure}

To compare the fit results with the distributions in data, four more moments, 
$\rm T1-T4$, are defined from Eq.~\eqref{formulation}:
\begin{equation}
    \begin{split}
        \rm{T1}^k &= \frac{1}{N_{\rm total}} \sum_{i}^{N^k} \left(\cos^2 \theta_{\Sigma} n^{i}_{1,z} n^{i}_{2,z} - \sin^2 \theta_{\Sigma} n^{i}_{1,x}n^{i}_{2,x} \right), \\ 
        \rm{T2}^k &= \frac{1}{N_{\rm total}} \sum_{i}^{N^k} \cos \theta_{\Sigma} \sin \theta_{\Sigma} \left( n^{i}_{1,z} n^{i}_{2,x} - n^{i}_{1,x}n^{i}_{2,z} \right),\\
        \rm{T3}^k &= \frac{1}{N_{\rm total}} \sum_{i}^{N^k} \cos \theta_{\Sigma} \sin \theta_{\Sigma} \left(n^{i}_{1,y} + n^{i}_{2,y} \right), \\
        \rm{T4}^k &= \frac{1}{N_{\rm total}} \sum_{i}^{N^k} \left(n^{i}_{1,z} n^{i}_{2,z} - \sin^2 \theta_{\Sigma} n^{i}_{1,y} n^{i}_{2,y} \right),
    \end{split}
\end{equation}
where $n_{1,x} = \cos \theta_{p} \sin \theta_{\Lambda} \cos
\varphi_{\Lambda}$, $n_{1,y} = \cos \theta_{p} \sin \theta_{\Lambda}
\sin \varphi_{\Lambda}$, $n_{1,z} = \cos \theta_{p} \cos
\theta_{\Lambda}$, $n_{2,x} = \cos \theta_{\bar{p}} \sin
\theta_{\bar{\Lambda}} \cos \varphi_{\bar{\Lambda}}$, $n_{2,y} = \cos
\theta_{\bar{p}} \sin \theta_{\bar{\Lambda}} \sin
\varphi_{\bar{\Lambda}}$, $n_{2,z} = \cos \theta_{\bar{p}} \cos
\theta_{\bar{\Lambda}}$.  Figure~\ref{T1_T4} shows the distributions
of these moments versus $\cos\theta_{\Sigma}$ for the $J/\psi$ dataset,
the ones for the $\psi(3686)$ dataset are shown 
in Ref.~\ref{T1_T4_3686}. 
The fit results are consistent with those distirbutions in data.

\begin{table*}[!htbp]
\renewcommand\arraystretch{1.2}
    \caption{Absolute systematic uncertainties ($\times 10^{-3}$) for
    the measured parameters.
    }
\label{tab:systematic_err}
\centering
\scalebox{1}
{
    \begin{tabular}{l c c c c c c c c | c c c}
        \hline \hline 
        \noalign{\vskip 3pt}
        Source ($10^{-3}$)                      & $\alpha_{J/\psi}$ & $\Delta\Phi_{J/\psi}$~(rad) & $\alpha_{\psi(3686)}$ & $\Delta\Phi_{\psi(3686)}$~(rad) & $\alpha_{\Sigma^0}$ & $\bar{\alpha}_{\Sigma^0}$ & $\alpha_{\Lambda}$ & $\bar{\alpha}_{\Lambda}$ & $A^{\Sigma}_{CP}$ & $A^{\Lambda}_{CP}$ \\
        \hline
        Tracking                                & 0.0 & 0.2 & 10 & 13 & 0.9 & 1.5 & 1 & 1 & 0.6 & 0 \\
        $\gamma$ reconstruction                 & 0.5 & 0.5 & 1  & 0  & 0.2 & 0.1 & 2 & 3 & 0.0 & 3 \\
        $\Lambda/\bar{\Lambda}$ reconstruction  & 6.2 & 0.2 & 24 & 30 & 0.7 & 0.4 & 6 & 3 & 0.2 & 8 \\
        4C kinematic fit                        & 1.5 & 0.1 & 5  & 6  & 1.0 & 1.5 & 2 & 3 & 0.6 & 3 \\
        $\Sigma^0/\bar{\Sigma}^0$ mass window   & 0.2 & 3.2 & 2  & 7  & 0.8 & 0.2 & 7 & 6 & 0.8 & 8 \\
        Background estimation                   & 4.3 & 0.4 & 10 & 6  & 0.6 & 0.1 & 6 & 6 & 0.6 & 9 \\
        \hline                                                                                        
        Total                                   & 7.7 & 3.3 & 28 & 34 & 1.8 & 2.2 &11 &10 & 1.3 & 15 \\ 
        \hline \hline
    \end{tabular}
    }
\end{table*}

The sources of the systematic uncertainties are summarized in
Table~\ref{tab:systematic_err}.  The total systematic uncertainties of
various parameters are obtained by summing the individual
contributions in quadrature.  The details are described as
follows. 

The control samples $J/\psi \to p \bar{p} \pi^+ \pi^-$, $e^+ e^- \to
\gamma \mu^+ \mu^-$, and $J/\psi \to pK^-\bar{\Lambda} + c.c.$ are
used to estimate the uncertainties from the $p,~\pi$ tracking,
$\gamma$ reconstruction, and $\Lambda/\bar{\Lambda}$
reconstruction~\cite{BESIII:2023olq}, respectively.  The efficiency
differences between data and MC simulation for the control samples are
used to re-weight the PHSP MC sample. The differences between the fit
results with corrections and the nominal fit are taken as the
systematic uncertainties.

For the 4C kinematic fit,  
some discrepancies are observed in the
$\chi^2_{\rm 4C}$ distributions between data and the signal MC sample.
Similar to Refs.~\cite{BESIII:2023ldd, BESIII:2024zav}, a data-driven
method is used to estimate these effects on the final results. 
The PHSP MC samples are weighted according to the $\chi^2_{\rm 4C}$
distributions in data, and the differences between the fit results
after weighting and the nominal results are taken as the systematic
uncertainties of the 4C kinematic fit.

The mass window of $\Sigma^0/\bar{\Sigma}^0$ is $\pm 3\sigma$ wide
around the known $\Sigma^0$ mass~\cite{ParticleDataGroup:2022pth},
where $\sigma = 4.5$ MeV/$c^2$ is the mass resolution of the
reconstructed $\Sigma^0/\bar{\Sigma}^0$. We change the mass window to
$\pm 2\sigma$ or $\pm 4\sigma$ to study the systematic uncertainties
from the $\Sigma^0/\bar{\Sigma}^0$ mass requirement. The largest
deviations from the nominal values are taken as the systematic
uncertainties.

The systematic uncertainties from the background estimation are
studied by including the background in the fit or not.  In the nominal
solution, the events in the data sideband region are used to subtract
the background events.  Here, an alternative solution is performed by
regarding all the selected events in the data sample as signal events.
The differences between the alternative and nominal fit results are
taken as the systematic uncertainties.

In summary, this Letter presents the first
measurements of the parity-violating decay parameters for the decays 
$\Sigma^0 \to \Lambda \gamma$ and $\bar{\Sigma}^0 \to \bar{\Lambda} \gamma$, 
$\alpha_{\Sigma^0} = -0.0017 \pm 0.0021 \pm 0.0018$ and
$\bar{\alpha}_{\Sigma^0} = 0.0021 \pm 0.0020 \pm 0.0022$,
respectively. The extracted values of decay parameters allow
for the strong-$CP$ test, 
$A^{\Sigma}_{CP} = (0.4 \pm 2.9 \pm 1.3)\times 10^{-3}$.
These results are consistent with $P$ and $CP$ conservation,
and will provide crucial constraints for new physics models. The
decay parameters $\alpha_{\Lambda}$ and $\bar{\alpha}_{\Lambda}$,
which are listed in Table~\ref{tab:final_results}, are measured
independently, and the weak-$CP$ test is performed in the sub-decays
of $\Lambda$ and $\bar{\Lambda}$.  These results are in good agreement
with the values obtained from the $J/\psi \to \Lambda \bar{\Lambda}$
and $J/\psi \to \Xi^{-(0)} \bar{\Xi}^{+(0)}$ analyses of
BESIII~\cite{BESIII:2021ypr, BESIII:2022qax, BESIII:2023drj}.

Furthermore, the spin polarizations of the $\Sigma^0$ hyperons with
opposite directions in the $J/\psi$ and $\psi(3686)$ decays are
observed for the first time. This phenomenon is also observed in the
case of the $\psi \to \Sigma^+ \bar{\Sigma}^-$ decays~\cite{BESIII:2020fqg}, 
but not in the $\Xi^{-(0)}$~\cite{BESIII:2023drj,BESIII:2023lkg,
BESIII:2021ypr, BESIII:2022lsz} hyperon pairs. Following
Ref.~\cite{Wu:2021yfv} with the measured parameters $\alpha_{\psi}$
and $\Delta \Phi_{\psi}$, the ratio of the D-wave and S-wave coupling
constants $g_D/g_S$, the relative phase $\delta$ between the S-wave
and D-wave, the percentage of the S-wave contribution
$\Gamma_S/\Gamma_{\rm total}$, and the effective radius $r_{\rm eff}$
of the $\psi\to \Sigma^0 \bar{\Sigma}^0$ decay 
(where $\bar{L} = r_{\rm eff} \times p$, $\bar{L}$ and $p$ 
are the average orbital angular momentum and the relative momentum 
of the $\Sigma^0$ and $\bar{\Sigma}^0$ in $\psi$ C.M. system) are obtained for the
first time, as shown in Table~\ref{S_D_wave} together with the
parameters for $\psi \to \Sigma^+ \bar{\Sigma}^-$ calculated in
Ref.~\cite{Wu:2021yfv}.  What is particularly intriguing is that, as
speculated in Ref.~\cite{Wu:2021yfv}, the $\delta$ difference between
$J/\psi \to \Sigma^0 \bar{\Sigma}^0$ and $\psi(3686) \to \Sigma^0
\bar{\Sigma}^0$ is about the same as the value between $J/\psi \to
\Sigma^+ \bar{\Sigma}^-$ and $\psi(3686) \to \Sigma^+ \bar{\Sigma}^-$
decays~\cite{Wu:2021yfv}, and both are equal to approximately $\pi$.
If this is not a coincidence, the underlying mechanism should be
investigated in the future.  These results are crucial to understand
the decay dynamics of the charmonium states and the production
mechanism of the $\Sigma^0-\bar{\Sigma}^0$ pairs and provide useful
information for hunting down excited nucleon
resonances~\cite{Wu:2021yfv}.

\begin{table*}[!htbp] \renewcommand\arraystretch{1.2}
    \caption{The parameters of the $\psi \to \Sigma^0 \bar{\Sigma}^0$ (this work) and 
    $\psi \to \Sigma^+ \bar{\Sigma}^-$ (from Ref.~\cite{Wu:2021yfv}) processes, 
    including the ratio between the coupling constants of D-wave and S-wave ($g_D/g_S$), 
    the relative phase between S-wave and D-wave ($\delta$), 
    the ratio between partial width of S-wave and total width ($\Gamma_S/\Gamma_{\rm total}$), 
    and the effective radius ($r_{\rm eff}$).
    The first and second uncertainties in this work are
    statistical and systematic, respectively, and the ones for $\psi \to \Sigma^+ \bar{\Sigma}^-$ are the total uncertainties.
    }
    \label{S_D_wave}
\centering
\scalebox{1}
{
    \begin{tabular}{l c c c c }
        \hline \hline 
        \noalign{\vskip 3pt}
        Mode                & $g_D/g_S$~(GeV$^{-2}$) & $\delta$~(rad) & $\Gamma_S/\Gamma_{\rm total}$~(\%) & $r_{\rm eff}$~(fm)\\
        \hline
        $J/\psi \to \Sigma^0 \bar{\Sigma}^0$&     $0.1217\pm 0.0015 \pm 0.0028$ & $2.947 \pm 0.017 \pm 0.012$ & $95.23\pm 0.11 \pm 0.21$ & $0.0191 \pm 0.0005 \pm 0.0008$ \\
        $\psi(3686) \to \Sigma^0 \bar{\Sigma}^0$& $0.123\pm 0.008 \pm 0.006$ & $-0.28 \pm 0.07 \pm 0.04$ & $82.7\pm 1.9 \pm 1.4$ & $0.049 \pm0.005 \pm 0.004$ \\
        $J/\psi \to \Sigma^+\bar{\Sigma}^-$~\cite{Wu:2021yfv}&     $0.171 \pm 0.006$ & $2.67 \pm 0.04$ & $90.9\pm 0.6$ & $0.0362 \pm 0.0024$ \\
        $\psi(3686) \to \Sigma^+ \bar{\Sigma}^-$~\cite{Wu:2021yfv}& $0.097 \pm 0.009$ & $-0.33 \pm 0.10$ & $88.3\pm 2.0$ & $0.033 \pm 0.006$  \\
        \hline \hline
    \end{tabular}
    }
\end{table*}

\textbf{Acknowledgement}

The BESIII Collaboration thanks the staff of BEPCII and the IHEP computing center for their strong support. The authors would like to extend thanks to Prof. Jusak Tandean, Dr. Shu-Ming Wu and Dr. Jia-Jun Wu for useful discussion and helpful advice. This work is supported in part by National Key R\&D Program of China under Contracts Nos. 2023YFA1606000, 2020YFA0406300, 2020YFA0406400; National Natural Science Foundation of China (NSFC) under Contracts Nos. 11635010, 11735014, 11935015, 11935016, 11935018, 12025502, 12035009, 12035013, 12061131003, 12165022, 12192260, 12192261, 12192262, 12192263, 12192264, 12192265, 12221005, 12225509, 12235017, 12342502, 12361141819; the Chinese Academy of Sciences (CAS) Large-Scale Scientific Facility Program; the CAS Center for Excellence in Particle Physics (CCEPP); Joint Large-Scale Scientific Facility Funds of the NSFC and CAS under Contract No. U1832207; 100 Talents Program of CAS; The Institute of Nuclear and Particle Physics (INPAC) and Shanghai Key Laboratory for Particle Physics and Cosmology; Yunnan Fundamental Research Project under Contract No. 202301AT070162; German Research Foundation DFG under Contracts Nos. 455635585, FOR5327, GRK 2149; Istituto Nazionale di Fisica Nucleare, Italy; Ministry of Development of Turkey under Contract No. DPT2006K-120470; National Research Foundation of Korea under Contract No. NRF-2022R1A2C1092335; National Science and Technology fund of Mongolia; National Science Research and Innovation Fund (NSRF) via the Program Management Unit for Human Resources \& Institutional Development, Research and Innovation of Thailand under Contract No. B16F640076; Polish National Science Centre under Contract No. 2019/35/O/ST2/02907; The Swedish Research Council; The Knut and Alice Wallenberg Foundation, Sweden; The Swedish Foundation for International Cooperation in Research and Higher Education (STINT); U. S. Department of Energy under Contract No. DE-FG02-05ER41374.

%

\end{document}